\documentclass[12pt,preprint]{aastex}

\begin{document}

\title{Can Gamma-Ray Bursts Be Used to Measure Cosmology?\\ A Further Analysis}

\author{D. Xu$^1$, Z. G. Dai$^1$, and E. W. Liang$^{1,2,3}$}

\affil{$^1$Department of Astronomy, Nanjing University, Nanjing 210093, China\\
$^2$Department of Physics, Guangxi University, Nanning 530004, China\\
$^3$Department of Physics, University of Nevada, Las Vegas 89154, USA}

\begin{abstract}
Three different methods of measuring cosmology with gamma-ray bursts (GRBs) have been proposed
since a relation between the $\gamma$-ray energy $E_{\gamma}$ of a GRB jet and the peak energy
$E_p$ of the $\nu F_{\nu}$ spectrum in the burst frame was reported by Ghirlanda and coauthors.
In Method I, to calculate the probability for a favored cosmology, only the contribution of the
$E_\gamma-E_p$ relation that is already best fitted for this cosmology is considered. We apply
this method to a sample of 17 GRBs, and obtain the mass density $\Omega_M=0.15^{+0.45}_{-0.13}$
($1\sigma$) for a flat $\Lambda$CDM universe. In Method II, to calculate the probability for
some certain cosmology, contributions of all the possible $E_\gamma-E_p$ relations that are
best fitted for their corresponding cosmologies are taken into account. With this method, we
find a constraint on the mass density $0.14<\Omega _M<0.69$ ($1\sigma$) for a flat universe. In
Method III, to obtain the probability for some cosmology, contributions of all the possible
$E_\gamma-E_p$ relations associated with their unequal weights are considered. With this
method, we obtain an inspiring constraint on the mass density $0.16<\Omega _M<0.45$ ($1\sigma$)
for a flat universe, and a $\chi^2_{dof}=19.08/15=1.27$ for the concordance model of
$\Omega_M=0.27$. Compared with the previous two methods, Method III makes the observed 17 GRBs
place much more stringent confidence intervals at the same confidence levels. Furthermore, we
perform a Monte Carlo simulation and use a larger sample to investigate the cosmographic
capabilities of GRBs with different methods. We find that, a larger GRB sample could be used to
effectively measure cosmology, no matter whether the $E_{\gamma}-E_p$ relation is calibrated by
low-$z$ bursts or not. Ongoing observations on GRBs in the \emph{Swift} era are expected to
make the cosmological utility of GRBs progress from its babyhood into childhood.
\end{abstract}


\section{Introduction}\label{sec:introduction}
The traditional cosmology has been revolutionized by modern sophisticated observation
techniques in distant Type Ia supernovae (SNe Ia) (e.g. Riess et al. 1998; Schmidt et al. 1998;
Perlmutter et al. 1999), cosmic microwave background (CMB) fluctuations (e.g. Bennett et al.
2003; Spergel et al. 2003), and large-scale structure (LSS) (e.g. Allen et al. 2003; Tegmark et
al. 2004). Each type of cosmological data trends to play an unique role in measuring cosmology.
In modern cosmology, it has been convincingly suggested that the global mass-energy budget of
the universe, and thus its dynamics, is dominated by a dark energy component, and that the
currently accelerating universe has once been decelerating (e.g., Riess et al. 2004). The
cosmography and the nature of dark energy as well as its evolution with redshift are one of the
most important issues in physics and astronomy today.

Gamma-ray bursts (GRBs) are the most intense explosions observed so far. They are believed to be
detectable up to a very high redshift (Lamb \& Reichart 2000; Ciardi \& Loeb 2000; Bromm \& Loeb
2002; Gou et al. 2004), and their high energy photons are almost immune to dust extinction. These
advantages would make GRBs an attractive cosmic probe.

From the isotropic-equivalent peak luminosity $L_{\rm iso}-$variability (or
spectral lag) relation (Fenimore \& Ramirez-Ruiz 2000; Norris, Marani, \&
Bonnell 2000), the standard energy reservior $E_{\gamma}$ of GRB jets (Frail et
al. 2001), the $L_{\rm iso}-$ peak energy $E_p$ of the $\nu F_{\nu}$ spectrum
in the burst frame relation (Lloyd-Ronning \& Petrosian 2002; Lloyd-Ronning \&
Ramirez-Ruiz 2002; Yonetoku et al. 2004), the isotropic-equivalent energy
$E_{\rm iso}-E_p$ relation (Amati et al. 2002) to the beaming-corrected energy
$E_\gamma-E_p$ relation (Ghirlanda et al. 2004a, Ghirlanda relation hereafter),
GRBs are towards more and more standardizable candles. However, these relations
in GRBs have not been calibrated by a low-$z$ GRB sample, so one should look
for a method which is different from the ``Classical Hubble Diagram'' method in
SNe Ia.

The luminosity relations with the variability and spectral lag make GRBs a distance indicator
in the same sense as Cepheids and SNe Ia, in which an observed light-curve property can yield
an apparent distance modulus (DM). Schaefer (2003) considered these two relations for nine
bursts with known redshifts and advocated a new cosmographic method (hereafter Method I) for
GRBs. In Method I, one first calibrates the two relations with the observed sample for a
certain cosmology, and then applies the best-fit relations back to the observed sample to
obtain a $\chi^2$ or a probability $P\propto {\rm exp(-\chi^2/2)}$ for this cosmology. Similar
to the brightness of SNe Ia, the energy reserviors in GRB jets are also clustered, but they are
not fine enough for precise cosmology (Bloom et al. 2003). Amati et al. (2002) found $E_{\rm
iso}\propto {E_p}^k$ ($k\sim 2$) relation from 12 \emph{BeppoSAX} bursts. The \emph{HETE-2}
observations confirm this relation and extend it to X-ray flashes (Sakamoto et al. 2004a; Lamb
et al. 2004). In addition, it also holds within a GRB (Liang, Dai \& Wu 2004). The Ghirlanda
relation is written as $(E_{\gamma}/10^{50 }{\rm ergs})=C ({E_p/100 \rm keV})^{a}$, where $a$
and $C$ are dimensionless parameters. Theoretical explanations of this relation include the
standard synchrotron mechanism in relativistic shocks (Zhang \& M\'esz\'aros 2002; Dai \& Lu
2002) together with the afterglow jet model, or the emission from off-axis relativistic jets
(Yamazaki, Ioka \& Nakamura 2004; Eichler \& Levinson 2004; Levinson \& Eichler 2005). This
relation could also be understood due to comptonization of the thermal radiation flux that is
advected from the base of an outflow in the dissipative photosphere model (e.g. Rees \&
M\'esz\'aros 2005). If these explanations are true, the Ghirlanda relation appears to be
intrinsic. Thus, Dai, Liang \& Xu (2004; DLX04) considered the Ghirlanda relation for 12 bursts
and proposed another cosmographic method (hereafter Method II) for GRBs. In Method II, one
makes marginalizations over the unknown parameters in the Ghirlanda relation to obtain a
$\chi^2$ or a probability $P\propto {\rm exp(-\chi^2/2)}$ for a certain cosmology. Following
Schaefer's method, Ghirlanda et al. (2004b; GGLF04) and Friedman \& Bloom (2004; FB04) also
investigated the same issue but used different GRB data. Recently, Firmani et al. (2005;
FGGA05) considered the Ghirlanda relation for 15 bursts and proposed a Bayesian approach for
the cosmological use (hereafter Method III). In Method III, to obtain the probability for a
certain cosmology, one considers contributions of all the possible $E_\gamma-E_p$ relations
associated with their unequal weights. The detailed procedures of the three methods are shown
in $\S2.1$, which indicate that Method III is the optimized one.

As analyzed previously, due to the lack of low-$z$ GRBs, Methods I, II and III are different
from the ``Classic Hubble Diagram'' method in SNe Ia. In this paper, we investigate the
constraints on cosmological parameters from the observed 17 GRBs with different methods.
Because the present GRB sample is a small one, it is necessary to use a large simulated sample,
which may be established in the \emph{Swift} era, to discuss the cosmographic capabilities with
different methods.

This paper is arranged as follows. In $\S2$, we describe our analytical methods and data. The
results from the observed GRB sample are presented in $\S3$. In $\S4$, we perform Monte Carlo
simulations and analyze the results from the simulated GRB sample. Conclusions and discussion
are presented in $\S5$.

\section{Method and Sample Analysis}\label{sec:analysis}
\subsection{Method Analysis}
According to the relativistic fireball model, the emission from a spherically expanding shell
and from a jet is similar to each other, if the observer is along the jet's axis and the
Lorentz factor of the fireball is larger than the inverse of the jet's half-opening angle
$\theta$; but when the Lorentz factor drops below $\theta ^{ - 1}$, the jet's afterglow light
curve is expected to present a break because of the edge effect and the laterally spreading
effect (Rhoads 1999; Sari, Piran \& Halpern 1999). Therefore, together with the assumptions of
the initial fireball emitting a constant fraction $\eta_\gamma$ of its kinetic energy into the
prompt $\gamma$-rays and a constant circumburst particle density $n$, the jet's half-opening
angle is derived to be
\begin{equation}
\theta  = 0.163\left( {\frac{{t_{j,d} }}{{1 + z}}} \right)^{3/8} \left( {\frac{{n_0 }}{{E_{{\rm
iso},52} }}\frac{{\eta _\gamma  }}{1 - \eta _\gamma}} \right)^{1/8},
\end{equation}
where $E_{{\rm iso},52}=E_{{\rm iso}}/10^{52}{\rm ergs}$, $t_{{\rm j},d}=t_{\rm j}/1\,{\rm
day}$, $n_0=n/1\,{\rm cm}^{-3}$. The ``bolometric'' isotropic-equivalent $\gamma$-ray energy of
a GRB is given by
\begin{equation}
E_{{\rm iso}}=\frac{4\pi d_L^2S_\gamma k}{1+z},
\end{equation}
where $S_\gamma$ is the fluence (in units of erg\,cm$^{-2}$) received in an observed bandpass and
the quantity $ k$ is a multiplicative correction of order unity relating the observed bandpass to a
standard rest-frame bandpass (1-$10^4$ keV in this paper) (Bloom, Frail \& Sari 2001). The energy
release of a GRB jet is thus given by
\begin{equation}
E_{\gamma}  = (1 - \cos \theta )E_{\rm iso},
\end{equation}
with the fractional uncertainty (FB04) being
\begin{eqnarray}
\left( {\frac{{\sigma _{E_\gamma  } }}{{E_\gamma  }}} \right)^2  & = & (1 - \sqrt {C_\theta  }
)^2 \left[ {\left( {\frac{{\sigma _{S_\gamma  } }}{{S_\gamma  }}} \right)^2  + \left(
{\frac{{\sigma _k }}{k}} \right)^2 } \right] +C_\theta \nonumber
     \\ & & \times \left[ {\left( {\frac{{3\sigma _{t_j } }}{{t_j }}} \right)^2
+ \left( {\frac{{\sigma _{n_0 } }}{{n_0 }}} \right)^2  + \left( {\frac{{\sigma _{\eta _\gamma }
}}{{\eta _\gamma- \eta^2 _\gamma  }}} \right)^2 } \right],
\end{eqnarray}
where
\begin{equation}
C_\theta   = \left[ {\theta \sin \theta /(8 - 8\cos \theta )} \right]^2.
\end{equation}
The Ghirlanda relation is
\begin{equation}
(E_{\gamma}/10^{50}{\rm ergs})=C(E_p/100\,{\rm keV})^{a},
\end{equation}
where $a$ and $C$ are assumed to have no covariance, and $E_p=E^{obs}_p(1+z)$. Combining Eqs.
(1), (2), (3) and (6), we derive the apparent luminosity distance with the small angle
approximation (i.e., $\theta\ll 1$)\footnote{This approximation is valid because $\left| ({1 -
\cos \theta - \theta ^2 /2})/(1 - \cos \theta) \right| <1\%$ when $\theta<0.35$ rad, and
$<0.4\%$ when $\theta<0.22$ rad.} as
\begin{equation}
d_L  = 7.575\frac{{(1 + z)C^{2/3} [E_p^{obs} (1 + z)/100\,{\rm keV}]^{2a/3} }}{{(kS_\gamma
t_{j,d} )^{1/2} (n_0 \eta _\gamma  )^{1/6} }}{\rm{ Mpc}},
\end{equation}
Assuming that all the observables are independent of each other and their errors satisfy
Gaussian distributions, we derive the fractional uncertainty of the apparent luminosity
distance without the small angle approximation,
\begin{eqnarray}
 \left( {\frac{{\sigma _{d_L } }}{{d_L }}} \right)^2  & = & \frac{1}{4}\left[
 {\left( {\frac{{\sigma _{S_\gamma  } }}{{S_\gamma  }}} \right)^2  + \left( {\frac{{\sigma _k
  }}{k}} \right)^2 } \right] +\frac{1}{4}\frac{{1 }}{{(1 - \sqrt
   {C_{\theta } } )^2 }}\left[ \left( {\frac{{\sigma _C }}{C}}\right)^2 \right. \nonumber \\ & &
   \left. + \left( {a\frac{{\sigma _{E_p^{obs} } }}{{E_p^{obs} }}} \right)^2 +
      \left( {a\frac{{\sigma _a }}{a}\ln \frac{E_p}{100}} \right)^2  \right]+ \frac{1}{4}\frac{C_{\theta }}{{(1 - \sqrt
   {C_{\theta } } )^2 }}  \nonumber \\ & &\times \left[ {\left( {\frac{{3\sigma _{t_j } }}{{t_j }}} \right)^2
    + \left( {\frac{{\sigma _{n_0 } }}{{n_0 }}} \right)^2}+ \left( {\frac{{\sigma _{\eta _\gamma  }
}}{{\eta _\gamma  (1 - \eta _\gamma  )}}} \right)^2 \right].
\end{eqnarray}
For simplicity, we consider $\eta_\gamma=0.2$ and $\sigma_{\eta_\gamma}=0$ throughout this
paper (Frail et al. 2001). The apparent DM of a burst can be given by
\begin{equation}
\mu_{\rm obs}= 5\log d_L + 25,
\end{equation}
with the uncertainty of
\begin{equation}
\sigma _{\mu _{{\rm obs}} } =\frac{5}{\ln 10} \frac{\sigma_{d_L}}{d_L}.
\end{equation}

On the other hand, the theoretical luminosity distance in $\Lambda$-models (Carroll, Press \&
Turner 1992) is given by
\begin{eqnarray}
d_L & = & c(1+z)H_0^{-1}|\Omega_k|^{-1/2}{\rm sinn}\{|\Omega_k|^{1/2}\nonumber  \\ & & \times
\int_0^zdz[(1+z)^2(1+\Omega_Mz)-z(2+z)\Omega_\Lambda]^{-1/2}\},
\end{eqnarray}
where $\Omega_k=1-\Omega_M-\Omega_\Lambda$, and ``sinn'' is $\sinh$ for $\Omega_k > 0$ and
$\sin$ for $\Omega_k < 0$. For $\Omega_k=0$, equation (11) degenerates to be $c(1+z)H_0^{-1}$
times the integral.

Usually, the likelihood for the parameters $\Omega_M$ and $\Omega_\Lambda$ can be determined
from a $\chi^2$ statistic, where
\begin{equation}
\chi ^2 (\Omega _M ,\Omega _\Lambda  ,a,C|h) = \sum\limits_{k} {\left[ {\frac{{\mu _{{\rm{th}}}
(z_k ;\Omega _M ,\Omega _\Lambda  |h) - \mu _{{\rm{obs}}} (z_k ;\Omega _M ,\Omega _\Lambda
,a,C|h)}}{{\sigma _{\mu _{{\rm{obs}}} (z_k ;\,\Omega _M ,\,\Omega _\Lambda,\,a,\,C,\,\sigma _a
/a,\,\sigma _C /C)} }}} \right]} ^2,
\end{equation}
where the dimensionless Hubble constant $h \equiv H_0 /100\,\,{\rm{ km\,s^{-1}Mpc^{-1}}}$ is
taken as 0.71. If the Ghirlanda relation could be calibrated by low-$z$ bursts, the above
$\chi^2$ statistic becomes the same as that in SNe Ia, that is,
\begin{equation}
\chi ^2 (\Omega _M ,\Omega _\Lambda  ,h) = \sum\limits_{k} {\left[ {\frac{{\mu _{{\rm{th}}}
(z_k ;\Omega _M ,\Omega _\Lambda  ,h) - \mu _{{\rm{obs,}}k} }}{{\sigma _{\mu _{{\rm{obs,}}k} }
}}} \right]} ^2,
\end{equation}
where $h$ should be marginalized.

The procedures of Method I , II and III are as follows:

\textbf{Method I} (see S03, GGLF04, FB04)

The procedure of this method is to (1) fix $\Omega _i  \equiv (\Omega _M ,\Omega _\Lambda )_i$,
(2) calculate $\mu_{\rm th}$ and $E_{\gamma}$ for each burst for that cosmology, (3) best fit
the $E_{\gamma}-E_p$ relation to yield $(a, C)_i$ and $(\sigma_a/a,\ \sigma_C/C)_i$, (4)
substitute $(a, C)_i$ and $(\sigma_a/a,\ \sigma_C/C)_i$ into Eqs. (7) and (8), and thus derive
$\mu_{\rm obs}$ and $\sigma _{\mu_{\rm obs}}$ for each burst for cosmology $\Omega _i$, (5)
calculate $\chi^2$ for cosmology $\Omega _i$ by comparing $\mu_{\rm th}$ with $\mu_{\rm obs}$,
$\sigma _{\mu_{\rm obs}}$, and then convert it to the probability by $P(\Omega _i)\propto \exp
( - \chi^2(\Omega _i) /2)$ (Riess et al. 1998), (6) repeat Steps 1$-$5 from $i=1$ to $i={\rm
N}$ to obtain the probability for each cosmology. Therefore, Method I is formulizd by
\begin{equation}
P(\Omega _i ) = P(\Omega _i |\Omega _i )\,\,\,\,\,\,\,\,\,\,  (i=1,\,{\rm N}).
\end{equation}

\textbf{Method II} (see DLX04)

The procedure of this method is to (1) fix $\Omega_i$, (2) calculate $\mu_{\rm th}$ and
$E_{\gamma}$ for each burst for that cosmology, (3)  best fit the $E_{\gamma}-E_p$ relation to
yield $(a, C)_i$ and $(\sigma_a/a,\ \sigma_C/C)_i$, (4) substitute $(a, C)_i$ and
$(\sigma_a/a,\ \sigma_C/C)_i$ into Eqs. (7) and (8), and thus derive $\mu_{\rm obs}$ and
$\sigma _{\mu_{\rm obs}}$ for each burst for cosmology $\Omega _i$, (5) repeat Steps 1$-$4 from
$i=1$ to $i={\rm N}$ to obtain $\mu_{\rm th}$, $\mu_{\rm obs}$ and $\sigma _{\mu_{\rm obs}}$
for each burst for each cosmology;\,\,\,\, (6) re-fix $\Omega_j$, (7) calculate
$\chi^2(\Omega_j |\Omega _i)$ by comparing $\mu_{\rm th}(\Omega_j)$ with $\mu_{\rm
obs}(\Omega_i)$, $\sigma _{\mu_{\rm obs}}(\Omega_i)$, and then convert it to a conditional
probability by $P(\Omega_j |\Omega _i)\propto \exp ( - \chi^2(\Omega_j |\Omega _i) /2)$, (8)
repeat Step 7 from $i=1$ to $i=\rm N$ to obtain the probability for cosmology $\Omega_j$ by
$P(\Omega _j ) \propto \sum\limits_i {\exp ( - \chi^2(\Omega_j |\Omega _i) /2)}$, (9) repeat
Steps 6$-$8 from $j=1$ to $j=\rm N$ to obtain the probability for each cosmology. Method II is
described by
\begin{equation}
P(\Omega _j ) = \sum\limits^{\rm N}_{i=1} {P(\Omega _j |\Omega _i )} \,\,\,\,\,\,(j=1,\,{\rm
N}).
\end{equation}

\textbf{Method III} (see FGGA05)

Method III is an improvement of Method II. Its key idea is to consider unequal weights for
different $E_\gamma-E_p$ relations, i.e., unequal weights for different conditional
probabilities $P(\Omega _j |\Omega _i )$. Therefore, the first seven steps of this method are
the same as those of Method II. The follow-up procedure is to (8) repeat Step 7 from $i=1$ to
$i=\rm N$ to obtain an iterative probability for cosmology $\Omega_j$ by $P^{\rm ite}(\Omega _j
) \propto \sum\limits_i {\exp ( - \chi^2(\Omega_j |\Omega _i) /2)\times P^{\rm ini}(\Omega _i
)}$ (here the initial probability $P^{\rm ini}(\Omega)$ for each cosmology is regarded as
equal; e.g. $P^{\rm ini}(\Omega)\equiv1$), (9) repeat Steps 6$-$8 from $j=1$ to $j=\rm N$ to
obtain an iterative probability $P^{\rm ite}(\Omega)$ for each cosmology; \,\,\,\,(10) replace
$P^{\rm ini}(\Omega)$ on Step 8 with $P^{\rm ite}(\Omega)$ on Step 9, then repeat Steps 8$-$9,
and thus reach another set of iterative probabilities for each cosmology, (11) run the above
cycle again and again until probability for each cosmology converges, i.e. $P^{\rm
ite}(\Omega)\Rightarrow P^{\rm fin}(\Omega)$ after tens of cycles.

In this method, to calculate the probability $P^{\rm fin}(\Omega_j)$ for a favored cosmology,
we consider contributions of all the possible $E_\gamma-E_p$ relations associated with their
weights. The conditional probability $P(\Omega_j |\Omega _i)$ denotes the contribution of some
certain relation, and $P^{\rm fin}(\Omega_i)$ weights the likelihood of this relation for its
corresponding cosmology. Therefore, the Bayesian approach can be formulized by
\begin{equation}
P^{\rm fin}(\Omega _j ) = \sum\limits^{\rm N}_{i=1} {P(\Omega _j |\Omega _i )\times P^{\rm
fin}(\Omega _i )} \,\,\,\,\,\,(j=1,{\rm N}).
\end{equation}

However, FGGA05 took different calculations for the conditional probability $P(\Omega_j |\Omega
_i)$. Making use of the incomplete gamma function, they transformed $\chi^2(\Omega_j |\Omega
_i)$ into its corresponding conditional probability $P(\Omega_j |\Omega _i)$. Actually, once
the parameters $(a, C)_i$ and $(\sigma_a/a,\ \sigma_C/C)_i$ of the Ghirlanda relation are
calibrated for cosmology $\Omega_i$, they become ``known'' for the cosmic model $\Omega_j$. So
herein the meaning of $\chi^2(\Omega_j |\Omega _i)$ is the same as that in SNe Ia. In this
paper, we redefine the conditional probability $P(\Omega_j |\Omega _i)$ by the formula of
$P(\Omega_j |\Omega _i)\propto \exp ( - \chi^2(\Omega_j |\Omega _i) /2)$.

\subsection{Sample Analysis}
The great diversity in GRB phenomena suggests that the GRB population may consist of
substantially different subclasses (e.g. MacFadyen \& Woosley 1999; Bloom et al. 2003; Sazonov
et al. 2004; Soderberg et al. 2004). To make GRBs a standard candle, a homogenous GRB sample is
required. The most prominent observational evidence for a GRB jet is its temporal break in
their afterglow light curves. For some bursts, e.g., GRB030329, their temporal breaks are
observed in both optical and radio bands. Berger et al. (2003) argued that these two breaks are
caused by the narrow component and wide component of the jet in this burst, respectively,
indicating that the physical origins of the breaks in the optical band and in the radio band
are different. In addition, the radio afterglow light curves fluctuate significantly. For
example, in the case of GRB970508, the light curve of its radio afterglow does not present
clearly a break. Only a lower limit of $t_j>25$ days was proposed by Frail et al. (2000).
Furthermore, the light curve of its optical afterglow is proportional to $t^{-1.1}$, in which
case no break appears (Galama et al. 1998). We thus include only those bursts whose temporal
breaks in optical afterglow light curves were well measured in our analysis. We obtain a sample
of 17 GRBs excluding GRB970508. They are listed in Table 1.

We correct the observed fluence in a given bandpass to a ``bolometric'' bandpass of $1-10^4 \
\rm{keV}$ with spectral parameters. The fluence and spectral parameters for a burst fitted by
different authors may be affected by different criterions (or systematical biases) in their
works. We thus collect a couple of fluence and spectral parameters from the same original
literature. For GRB 970828, GRB 980703, GRB 991216, and GRB 011211, the fluence and spectral
parameters are unavailable in the same original literature, so we choose their fluences
measured in the widest energy band available in the other literature.


For GRB 011211, we approximately take the high-energy spectral index $\beta$ to be $-2.3$
because it is unreported in Amati (2004). The spectra of {\em HETE-2}-detected GRB020124,
GRB020813, GRB021004, GRB030226 and XRF030429 are not fitted by the Band function but the
cutoff power law model. However, it is appropriate that their corresponding ``bolometric''
fluences are calculated by the former model with $\beta\sim-2.3$, avoiding the potential
systematical bias which is brought by applying different spectral models for one observed
sample (Barraud et al. 2003).

The circumburst densities of several bursts in our sample have been obtained from broadband
modelling of the afterglow emission (e.g., Panaitescu \& Kumar 2002). For the bursts with
unknown $n$, we assumed $n\simeq3 \,{\rm cm}^{-3}$ as the median value of the distribution of
the measured densities, together with a constant fractional uncertainty of $80\%$ (Ghirlanda et
al. 2004a; DLX04).

\section{Cosmological Constraints}
We first fit the $E_\gamma-E_p$ relation for the cosmology of $\Omega_M=0.27$,
$\Omega_\Lambda=0.73$, and $h=0.71$, and obtain $a=1.53$, $C=0.97$, $\sigma_a/a=0.05$ and
$\sigma_C/C=0.08$, together with the $\chi^2_{\nu}=21.93/15.0=1.46$, using Eq. (4) for the
estimation of $\sigma_{E_\gamma}/E_\gamma$ (Press et al. 1999). Substituting the best-fit
results into Eqs. (7) and (8), we plot the Hubble diagram for the observed GRB sample in the
concordance model of $\Omega_M=0.27$, which is shown in Fig 1 (filled circles). Comparatively
shown are the Hubble diagrams for the binned gold sample of SNe Ia (open circles; Riess et al.
2004), and the theoretical model of $\Omega_M=0.27$ and $\Omega_\Lambda=0.73$ (solid line).

The constraints with Method I are shown in Fig 2 (solid contours). In the concordance model of
$\Omega_M=0.27$, we find a $\chi^2_{dof}=17.91/15=1.19$. We measure
$\Omega_M=0.15^{+0.45}_{-0.13}$ ($1\sigma$) for a flat $\Lambda$CDM universe. DLX04 proposed
Method II based on the principle that if there are unknown cosmology-independent parameters in
the $\chi^2$ statistic, they are usually marginalized over (i.e. integrating the parameters
according to their probability distribution). Note that DLX04 let the parameter $a$ of the
Ghirlanda relation intrinsically equal to 1.50. For the purpose of universality, in this paper
we let it vary freely, similar to the parameter $C$ of this relation (see details in $\S2.1$).
The constraints with Method II are shown in Fig 3 (solid contours). This method provides a more
stringent constraint of $0.14<\Omega _M<0.69$ ($1\sigma$) for a flat universe.

By Method I, we recalibrate the Ghirlanda relation for each cosmology, i.e., $\Omega_M$ and
$\Omega_\Lambda$ taken from 0 to 1. We find the half-opening angles of all the bursts are less
than 0.23 rad, and that the mean $\sigma_a/a$ and $\sigma_C/C$ are 0.049 and 0.083. As a
result, the typical error terms in Eq. (8) are $\sigma _{S_\gamma  } /2S_\gamma \sim 0.052$,
$\sigma _k /2k \sim 0.025$, $[\sqrt {C_\theta } /(2 - 2\sqrt {C_\theta  } )](3\sigma _{t_j }
/t_j ) \sim 0.091$, $[\sqrt {C_\theta } /(2 - 2\sqrt {C_\theta  } )](\sigma _{n_0 } /n_0 ) \sim
0.118$, $[ 1 /(2 - 2\sqrt {C_\theta } )](\sigma _C /C ) \sim 0.055$, $a[ 1 /(2 - 2\sqrt
{C_\theta } )](\sigma _{E_p^{obs} } /E_p^{obs}) \sim 0.168$, and $\sigma_a[ 1 /(2 - 2\sqrt
{C_\theta  } )]\left| {\ln [E_p^{obs} (1 + z)/100]} \right|\sim 0.056$. These error terms give
a typical uncertainty of apparent DM $\sigma _{\mu _{{\rm{obs}}}}\sim0.5$ magnitudes, which is
a factor of $\sim2$ larger than that derived from the SN Ia gold sample.

Firmani et al. (2005) proposed the Bayesian approach to use GRBs as cosmic rulers. Being
different from their work, we redefined the conditional probability in their method and called
it Method III in this paper. The constraints on cosmological parameters with this method are
shown in Fig 4 (solid contours). Compared with previous two methods, Method III does give much
more stringent confidence intervals at the same confidence levels (C.L.). The results are
inspiring and demonstrate the advantages of high-$z$ distance indicators in constraining
cosmological parameters. The data set is consistent with the concordance model of
$\Omega_M=0.27$, yielding a $\chi^2_{dof}=19.08/15=1.27$. We also find $0.16<\Omega _M<0.45$
($1\sigma$) for a flat universe.

\section{Simulations and Cosmological Constraints}\label{simu}
\subsection{Procedure of Simulations}\label{procedure}
As discussed above, the method for the GRB cosmology is different from that for the SN
cosmology due to the lack of the low-$z$ calibration. The low-$z$ calibration shall greatly
enhance the cosmographic capability of GRBs (assuming no cosmic evolution for the Ghirlanda
relation). Therefore, one may ask: \emph{could a large high- $z$ GRB sample effectively measure
cosmology?} To answer this question, we carry out a Monte Carlo simulation and use a large
simulated sample to investigate the cosmographic capabilities in different scenarios.

Our simulations are based on the Ghirlanda relation which is calibrated for $\Omega_M=0.27$ and
$\Omega_\Lambda=0.73$. Making use of the sample in Table 1, we find $a=1.53$, $C=0.97$,
$\theta<0.2$ rad, and $\log(E_{\gamma}/1\,{\rm erg})\in [49.82,51.96]$. We also find
$\log(E^{obs}_p/1\,{\rm KeV})\in[1.54,2.89]$. These restrictive conditions are imposed upon our
simulations. Each simulated GRB is characterized by a set of $S'_b$, $t'_j$, $n'$, $z$ and
$E'_p$, where $S'_b$ is the bolometric fluence.

The simulation procedure is as follows:

(1) We consider the lognormal distributions for the observables $S_b$, $t_j$ and $n$. From the
17 GRBs, we find $\left\langle {\log \left( {S_b /1\,\rm{erg}\,{\rm cm}^{-2} } \right)}
\right\rangle  \sim  - 4.46$ with $\sigma_{\log \left({S_b /1\,\rm{erg}\,{\rm
cm}^{-2}}\right)}\sim0.78$, and $\left\langle {\log \left( {t_j /1\,{\rm{ day}}} \right)}
\right\rangle  \sim 0.17$ with $\sigma _{\log \left( {t_j /1\,{\rm{day}}} \right)}  \sim 0.27$.
Because $n$ is unavailable in the literature for most GRBs, we take $\left\langle {\log \left(
{n/1\,{\rm{ cm}}^{ - 3} } \right)} \right\rangle  \sim 0.40$ with $\sigma _{\log \left(
{n/1\,{\rm{cm}}^{ - 3} } \right)}  \sim 0.25$. The observable $z$ is selected according to its
observational distribution cut by an upper limit $z\sim4.5$, which is shown in Fig 5. The
uncertainty of $z$ is ignored.

(2) We also consider the lognormal distributions for the fractional uncertainties of the
observables $S_b$, $t_j$, $n$ and $E_p$. From the observed sample, we find $\left\langle {\log
\left( {\sigma _{S_b } /S_b } \right)} \right\rangle  \sim  - 0.97$ with $\sigma _{\log [
{\sigma _{S_b } /S_b } ]}  \sim 0.16$, $\left\langle {\log \left( {\sigma _{t_j } /t_j }
\right)} \right\rangle  \sim  - 0.91$ with $\sigma _{\log [ {\sigma _{t_j } /t_j } ]}  \sim
0.45$, $\left\langle {\log \left( {\sigma _n /n} \right)} \right\rangle  \sim  - 0.30$ with
$\sigma _{\log [ {\sigma _n /n} ]}  \sim 0.10$, and $\left\langle {\log \left( {\sigma _{E_p }
/E_p } \right)} \right\rangle  \sim  - 0.83$ with $\sigma _{\log [ {\sigma _{E_p } /E_p } ]}
\sim 0.26$, respectively. We ensure in code that the fractional uncertainties of the
observables $S_b$, $t_j$, $n$ and $E_p$ are less than $25\%$, $35\%$, $100\%$ and $35\%$,
respectively.

(3) We simulate a GRB characterized by a set of ($S_b\pm \sigma_{S_b}$, $t_j\pm \sigma_{t_j}$,
$n\pm \sigma_{n}$ and $z$) according to the distributions that these parameters follow, compute
its $E_\gamma$ in the cosmic model of $\Omega_M=0.27$ and $\Omega_\Lambda=0.73$, and then
calculate its $E_p$ by the Ghirlanda relation of $(E_{\gamma}/10^{50 }{\rm ergs})=0.97
({E_p/100 \rm keV})^{1.53}$.

(4) The GRB generated in step (3) follows the ``rigid'' Ghirlanda relation. We add a random
deviation to each parameter, except for $z$, to make this burst more realistic, i.e.,
$S'_b=S_b+1.1(-1)^{m} \sigma_{S_b}$, $t'_j=t_j+1.1(-1)^{m} \sigma_{t_j}$, $n'=n+1.1(-1)^{m}
\sigma_{n}$ and $E_p'=E_p+1.1(-1)^m \sigma_{E_p}$, where $m$ is randomly taken from 0 and 1.

(5) Using the parameters $S'_b$, $t'_j$, $n'$, $z$, and $E_p'$, we compute the parameters
$\theta'$ and $E'_\gamma$ for $\Omega_M=0.27$ and $\Omega_\Lambda=0.73$, and the quantity
$E'^{\,obs}_p=E'_p/(1+z)$.

(6) Since $\theta<0.2$ rad, $\log(E_{\gamma}/1\,{\rm erg})\in [49.82,51.96]$ and
$\log(E^{obs}_p/1\,{\rm KeV})\in[1.54,2.89]$ are valid for the observed sample, we require that
$\theta'$, $\log(E'_{\gamma}/1\,{\rm erg})$, and $\log(E'^{\,obs}_p/1\,{\rm KeV})$ of a
simulated GRB must be within the corresponding ranges.

(7) Repeat steps (3)-(6) to generate a sample of 80 bursts.

The half-opening angles of the simulated sample are less than 0.23 rad when $\Omega_M$ and
$\Omega_\Lambda$ are taken from 0 to 1, as in the observed sample. We carry out a circular
operation to achieve the simulated sample, which may be established in the \emph{Swift} era.

\subsection{Constraints from Simulated GRBs}\label{meth}
By Method I, we find the typical $\sigma_a/a$ and $\sigma_C/C$ for the large sample decrease to
0.02 and 0.05. This is because for the small observed sample, the dispersion of the Ghirlanda
relation is mainly contributed by a few ``outliers'', while for the large simulated sample, the
bursts are distributed around the ``rigid'' Ghirlanda relation with a Gaussian distribution
(see $\S4.1$). Such a large sample seems to be more realistic.

Among several scenarios, we only perform to what degree the simulated sample can constrain the
$\Omega_M-\Omega_\Lambda$ parameters with Method III. The results are shown by solid confidence
intervals in Fig 6. Also shown are the constraints derived from the SN gold sample (dashed
contours) at the same confidence levels. The main points revealed by this figure are: (1) A
large high-$z$ GRB sample could effectively measure cosmology. The simulated GRB data are
consistent with the concordance model of $\Omega_M\approx0.3$, yielding a
$\chi^2_{dof}=94./78.\approx1.20$. With the prior of a flat universe, the mass density is
$\Omega_M=0.30^{+0.09}_{-0.06}$ at the $68.3\%$ level. (2) GRBs can well constrain $\Omega_M$
rather than $\Omega_\Lambda$ due to their high redshifts. The orientation of the elliptical
contours is almost vertical to the $\Omega_M$ axis. This is an advantage of GRBs over SNe Ia
for the cosmological use. (3) GRBs supply complemental content to the SN cosmology. The 157 SNe
Ia provide evidence of cosmic acceleration at a very high confidence level. A large GRB sample
would reach a similar conclusion. Alone the SN sample nearly rules out the flat universe model
at $1\sigma$ level, but a combination of GRBs and SNe makes the concordance model of
$\Omega_M\approx0.3$ more favored and thus more agreement with the conclusion from WMAP
observations (Spergel et al. 2003).

If the low-$z$ calibration is realized for GRBs, then the solid confidence regions in Fig 6
will become smaller so that they lie in the part of cosmic acceleration at a high C.L. in the
$\Omega_M-\Omega_\Lambda$ plane (not shown in this work). We here present a rough estimation of
the detection rate of low-$z$ bursts in the \emph{Swift} era. Taking $\Omega_M=0.27$,
$\Omega_\Lambda=0.73$, and $h=0.71$, the observed rate of bursts with redshift less than $z$ is
\begin{equation}
\frac{{dN}}{{dt}} = \int_0^z {dz\frac{{dV(z)}}{{dz}}} \frac{{R_{GRB} (z)}}{{1 + z}},
\end{equation}
where $R_{GRB} (z)$ is the comoving GRB rate density, and $dV(z)/dz$ is the isotropic comoving
volume element,
\begin{equation}
\frac{{dV}}{{dz}} = 4\pi \left[ {\int_0^z {\frac{{dz'}}{{\sqrt {\Omega _M (1 + z')^3  + \Omega
_\Lambda  } }}} } \right]^2 \frac{1}{{\sqrt {\Omega _M (1 + z)^3  + \Omega _\Lambda  } }}.
\end{equation}
We assume $R_{GRB} (z) \propto R_{SN} (z) \propto R_{SF} (z)$, where $R_{SN} (z)$ and $R_{SF}
(z)$ are the comoving rate densities of core-collapse supernovae and star formation,
respectively (Porciani \& Madau 2001). We also assume the constant fraction
$k=R_{GRB}/R_{SN}\sim10^{-5}$, which considers the GRB formation efficiency out of the
core-collapse SNe and the GRB beaming effect. Additionally, we take $R_{SN} (z)\simeq0.01R_{SF}
(z)\,{\rm M^{-1}_{\bigodot}}$. The global star formation rate for the Einstein-de Sitter
universe (Steidel et al. 1999) is
\begin{equation}
R_{SF} (z) = 0.16\frac{{\exp (3.4z)}}{{\exp (3.4z) + 22}}\,\,\,{\rm{M_{\bigodot}\,yr}}^{{\rm{ -
1}}} {\rm{Mpc}}^{{\rm{ - 3}}}.
\end{equation}
Thus, we derive the detective probability of low-$z$ GRBs in the \emph{Swift} era (2$-$4
years): when $z \le 0.1$, there will be $<1$ burst; and when $z \le 0.2$, there will be a few
bursts. These results agree with the present observational GRB data. In addition, taking into
account possible different origins of very low-$z$ bursts (e.g. GRB980425) and high-$z$ bursts,
it might not be valid to directly apply the low-$z$ calibrated relation to a high-$z$ sample.
However, if the cosmic evolution of the ``standard candle'' relation is ignored, then in the
future one could consider a sample of GRBs with redshift $z \le 0.2$ for the low-$z$
calibration. As the zero-order approximation, the theoretical luminosity distance of such a GRB
sample can be written as $d_L (z) = z(c/H_0)$.

\section{Conclusions and Discussion}\label{discussion}

At present, GRBs with known redshifts are about 45 out of a few thousands, among which 17 are
available to derive the Ghirlanda relation. This relation is so tight that it has been
considered for the cosmological use. Although the low-$z$ calibration of the Ghirlanda relation
is not realized, the observed 17 high-$z$ bursts still provide interesting and even inspiring
results. We find GRBs independently place a constraint on the mass density $0.16<\Omega
_M<0.45$ ($1\sigma$) for a flat $\Lambda$CDM model.

GRBs are becoming more and more standardized candles. Through the Ghirlanda relation, the mean
scatter in GRBs is a factor of $\sim2$ larger than that in SNe Ia. However, the disadvantage of
larger scatter in GRBs is compensated, to some extent, by their advantages of high redshifts
and immunity to the dust extinction. The shape of the constraints in Fig 6 implies that GRBs
could not only measure the mass density $\Omega_M$ but also provide complementarity to the SN
cosmology. The {\it Swift} satellite is hopeful to establish a large GRB sample with known
redshifts, perhaps including low-$z$ bursts. In this sense, the GRB cosmology now lies in its
babyhood.

A reliable theoretical basis of the Ghirlanda relation is also important for GRBs as a cosmic
ruler. Using the small angle approximation for all bursts, a scaling analysis gives
$E_{\gamma}\propto{E^a_p}$ ($a\sim1.5$) as long as the parameters such as the spectral index
$p$ of the distribution of accelerated electrons, the the energy equipartition factor
$\varepsilon _e$ of the electrons, the energy equipartition factor $\varepsilon _B$ of magnetic
field, the bulk Lorentz factor $\gamma$, etc., or their combinations are clustered (Zhang \&
M\'esz\'aros 2002; Dai \& Lu 2002; Wu, Dai \& Liang 2004). This power-law relation could also
be understood by the emissions from off-axis relativistic jets (Yamazaki, Ioka \& Nakamura
2004; Eichler \& Levinson 2004; Levinson \& Eichler 2005) and the dissipative photosphere model
(Rees \& M\'esz\'aros 2005). Different plausible explanations imply that this topic needs
further investigations.

It should be pointed out that the Ghirlanda relation is obtained under the
framework of the uniform top-hat jet model. Other input assumptions include the
uniform circumburst medium density $n$ and the constant efficiency
$\eta_\gamma$ of converting the initial ejecta's kinetic energy into
$\gamma$-ray energy release. However, $n$ and $\eta_\gamma$ should be different
from burst to burst, and $n$ is variable for a burst in the wind environment
(Dai \& Lu 1998; Chevalier \& Li 1999). Thus, the quantity $n\eta_\gamma$ in
the Ghirlanda relation might not be clustered for the observed bursts (Friedman
\& Bloom 2004). New relations with as few well-observed quantities as possible
are required for improvement\footnote{After the submission of this paper, Liang
\& Zhang (2005) and Xu (2005) proposed new relations between the isotropic
$\gamma$-ray energy and the $\nu F_\nu$ peak energy by considering the break
time of an afterglow light curve. Clearly, these three quantities are directly
observed.}. In this paper, for those bursts with unknown $n$, we assumed
$n\simeq3 \,{\rm cm}^{-3}$ as the median value of the distribution of the
measured densities, together with a constant fractional uncertainty of $80\%$
(Ghirlanda et al. 2004a; DLX04). We also treat $\eta_\gamma=0.2$ and
$\sigma_{\eta_\gamma}=0$ for all the 17 bursts.

Finally, the Ghirlanda relation is not valid for all the observed BATSE sample.
This implies that this relation may suffer from the data selection effect (Band
\& Preece 2005). Its validity will be tested by the ongoing observations of the
\emph{Swift} mission. However, no matter whether the low-$z$ sample is
established or not, the GRB cosmology is expected to progress much in the
coming years.

\acknowledgments XD is grateful to X.L. Luo for helpful discussions. We thank
the anonymous referees for valuable comments that have allowed us to improve
our manuscript significantly. This work was supported by the National Natural
Science Foundation of China (grants 10233010, 10221001, and 10463001), the
Ministry of Science and Technology of China (NKBRSF G19990754), the National
Post Doctoral Foundation of China, and the Research Foundation of Guangxi
University.

\clearpage
\begin{deluxetable}{lcccccccccc}
\rotate \tabletypesize{\scriptsize} \tablewidth{8.2in}

\tablecaption{\label{Table1}Sample of 17 GRBs} \tablecolumns{11} \tablehead{
\colhead{GRB}      & \colhead{$z$}    & \colhead{$E_p^{obs}(\sigma_{E_p}^{obs})$}     &\colhead{$[\alpha, \beta]$}       & \colhead{$S_\gamma(\sigma_{S_\gamma})$}   & \colhead{bandpass}     & \colhead{$t_{\rm j}(\sigma_{t_{\rm j}})$}     & \colhead{$n (\sigma_n)$}  & \colhead{$\eta_\gamma$}  & \colhead{References}  & \colhead{Detected}\\
\colhead{}      & \colhead{}        &\colhead{[KeV]}     &\colhead{}     & \colhead{[$10^{-6}$ erg cm$^{-2}$]}       & \colhead{[KeV]}       & \colhead{days}   & \colhead{[cm$^{-3}$]} & \colhead{} & \colhead{($z$, $E_p^{obs}$, $[\alpha, \beta]$, $S_{\gamma}$, $t_{\rm jet}$, $n$)} & \colhead{by}\\
\colhead{}      & \colhead{}   &\colhead{\tablenotemark{a}}   &\colhead{\tablenotemark{a}}
&\colhead{\tablenotemark{b}}   &\colhead{\tablenotemark{b}}   &\colhead{\tablenotemark{c}}
&\colhead{\tablenotemark{d}}   &\colhead{\tablenotemark{e}}   &\colhead{\tablenotemark{f}}
&\colhead{}}\startdata
grb970828   &  0.9578  &   297.7[59.5]    &  -0.70, -2.07   &  96.0[9.6]      & 20-2000   & 2.2(0.4)     & 3.0[2.4]     &0.2   &01,18,18,26,01,no     &BATSE\\
grb980703   &  0.966   &   254.0[50.8]    &  -1.31, -2.40   &  22.6[2.26]     & 20-2000   & 3.4(0.5)     & 28.0(10.0)   &0.2   &02,18,18,26,27,27     &BATSE\\
grb990123   &  1.600   &   780.8(61.9)    &  -0.89, -2.45   &  300.0(40.0)    & 40-700    & 2.04(0.46)   & 3.0[2.4]     &0.2   &03,19,19,19,28,no     &SAX\\
grb990510   &  1.619   &   161.5(16.0)    &  -1.23, -2.70   &  19.0(2.0)      & 40-700    & 1.57(0.03)   & 0.29(0.14)   &0.2   &04,19,19,19,29,41     &SAX\\
grb990705   &  0.8424  &   188.8(15.2)    &  -1.05, -2.20   &  75.0(8.0)      & 40-700    & 1.0(0.2)     & 3.0[2.4]     &0.2   &05,19,19,19,30,no     &SAX\\
grb990712   &  0.4331  &   65.0(10.5)     &  -1.88, -2.48   &  6.5(0.3)       & 40-700    & 1.6(0.2)     & 3.0[2.4]     &0.2   &06,19,19,19,31,no     &SAX\\
grb991216   &  1.020   &   317.3[63.4]    &  -1.23, -2.18   &  194.0[19.4]    & 20-2000   & 1.2(0.4)     & 4.7(2.8)     &0.2   &07,18,18,26,32,41     &BATSE\\
grb011211   &  2.140   &   59.2(7.6)      &  -0.84, -2.30   &  5.0[0.5]       & 40-700    & 1.56(0.02)   & 3.0[2.4]     &0.2   &08,20,20,26,33,no     &SAX\\
grb020124   &  3.200   &   120.0(22.6)    &  -1.10, -2.30   &  6.8[0.68]      & 30-400    & 3.0(0.4)     & 3.0[2.4]     &0.2   &09,21,21,21,34,no     &HETE-2\\
grb020405   &  0.690   &   192.5(53.8)    &   0.00, -1.87   &  74.0(0.7)      & 15-2000   & 1.67(0.52)   & 3.0[2.4]     &0.2   &10,22,22,22,22,no     &SAX\\
grb020813   &  1.255   &   212.0(42.0)    &  -1.05, -2.30   &  102.0[10.2]    & 30-400    & 0.43(0.06)   & 3.0[2.4]     &0.2   &11,21,21,21,35,no     &HETE-2\\
grb021004   &  2.332   &   79.8(30.0)     &  -1.01, -2.30   &  2.55(0.60)     & 2 -400    & 4.74(0.14)   & 30.0(27.0)   &0.2   &12,23,23,23,36,42     &HETE-2\\
grb021211   &  1.006   &   46.8(5.5)      &  -0.805,-2.37   &  2.17(0.15)     & 30-400    & 1.4(0.5)     & 3.0[2.4]     &0.2   &13,24,24,24,37,no     &HETE-2\\
grb030226   &  1.986   &   97.1(20.0)     &  -0.89, -2.30   &  5.61(0.65)     & 2 -400    & 1.04(0.12)   & 3.0[2.4]     &0.2   &14,23,23,23,38,no     &HETE-2\\
grb030328   &  1.520   &   126.3(13.5)    &  -1.14, -2.09   &  36.95(1.40)    & 2 -400    & 0.8(0.1)     & 3.0[2.4]     &0.2   &15,23,23,23,39,no     &HETE-2\\
grb030329   &  0.1685  &   67.9(2.2)      &  -1.26, -2.28   &  110.0(10.0)    & 30-400    & 0.48(0.03)   & 1.0(0.11)    &0.2   &16,25,25,25,40,43     &HETE-2\\
xrf030429   &  2.658   &   35.0(9.0)      &  -1.12, -2.30   &  0.854(0.14)    & 2 -400    & 1.77(1.0)    & 3.0[2.4]     &0.2   &17,23,23,23,17,no     &HETE-2\\
\enddata

\tablenotetext{a} {The spectral parameters fitted by the Band function. The fractional
uncertainties of $E_p^{obs}$ are taken as $20\%$ when not reported, and the fractional
uncertainty of $k$-correction is fixed as $5\%$.}

\tablenotetext{b} {The fluences and their errors in the observed energy band. The fractional
errors are taken as $10\%$ when not reported. The fluence and spectral parameters of a GRB are
selected from the same original literature as possible. If this criterion unsatisfied, fluence
are chosen in the widest energy band.}

\tablenotetext{c} {Afterglow break times and errors in the optical band.}

\tablenotetext{d} {The circumburst densities and errors from broadband modelling of the
afterglow light curves. If no available the value of $n$ is taken as $3.0\pm2.4$ cm $^{-3}$.}

\tablenotetext{e} {The constant efficiency $\eta_\gamma$ of converting explosion energy into
$\gamma$-ray emission for each burst.}

\tablenotetext{f} {References in order for $z$, $E_p^{obs}$, $[\alpha, \beta]$, $S_{\gamma}$,
$t_{\rm j}$, and $n$.}

\tablerefs{ (1) Djorgovski et al. 2001; (2) Djorgovski et al. 1998; (3) Kulkarni et al. 1999; (4)
Vreeswijk et al. 2001; (5) Le Floc'h et al. 2002; (6) Vreeswijk et al. 2001; (7) Piro et al. 2000;
(8) Holland et al. 2002; (9) Hjorth et al. 2003; (10) Price et al. 2003a; (11) Barth et al. 2003;
(12) Matheson et al. 2003; (13) Vreeswijk et al. 2003; (14) Greiner et al. 2003a; (15) Rol et al.
2003; (16) Greiner et al. 2003b; (17) Jakobsson, et al. 2004; (18) Jimenez et al. 2001; (19) Amati
et al. 2002; (20) Amati 2004; (21) Barraud et al. 2003; (22) Price et al. 2003a; (23) Sakamoto et
al. 2004b; (24) Crew et al. 2003; (25) Vanderspek et al. 2004; (26) Bloom et al .2003; (27) Frail
et al. 2003; (28) Kulkarni et al. 1999; (29) Stanek et al. 1999; (30) Masetti et al. 2000; (31)
Bjornsson et al. 2001; (32) Halpern et al. 2000; (33) Jakobsson et al. 2003; (34) Berger et al.
2002; (35) Barth et al. 2003; (36) Holland et al. 2003; (37) Holland et al. 2004; (38) Klose et al.
2004; (39) Andersen et al. 2003; (40) Price et al. 2003b; (41) Panaitescu \& Kumar 2002; (42)
Schaefer et al. 2003; (43) Tiengo et al. 2003 }

\end{deluxetable}

\clearpage

\begin{figure}
\plotone{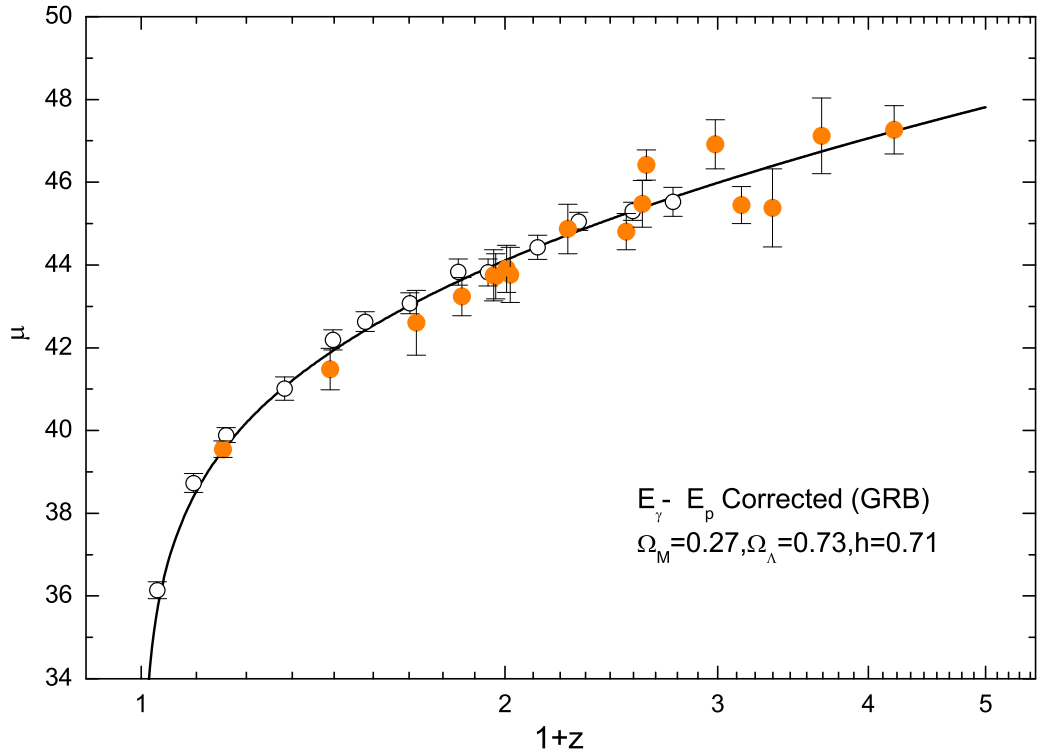} \caption{Hubble diagrams for the observed GRB sample (filled circles) and
for the binned gold sample of SNe Ia (open circles). The GRB $E_\gamma-E_p$ relation has been
calibrated in the cosmic model of $\Omega_M=0.27$, $\Omega_\Lambda=0.73$ and $h=0.71$ (solid
line). \label{fig1}}
\end{figure}

\begin{figure}
\plotone{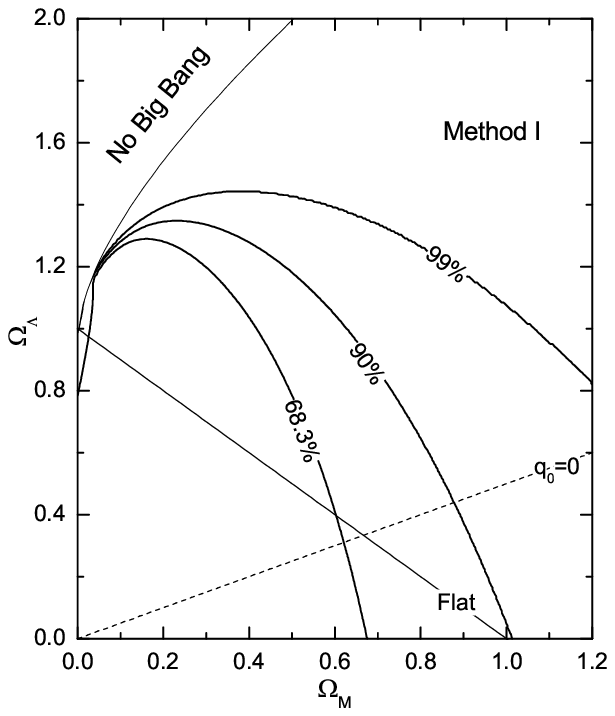} \caption{Joint confidence intervals (68.3\%, 90\% and 99\%) in the
$\Omega_M-\Omega_\Lambda$ plane from the 17 GRBs with Method I.\label{fig2}}
\end{figure}

\begin{figure}
\plotone{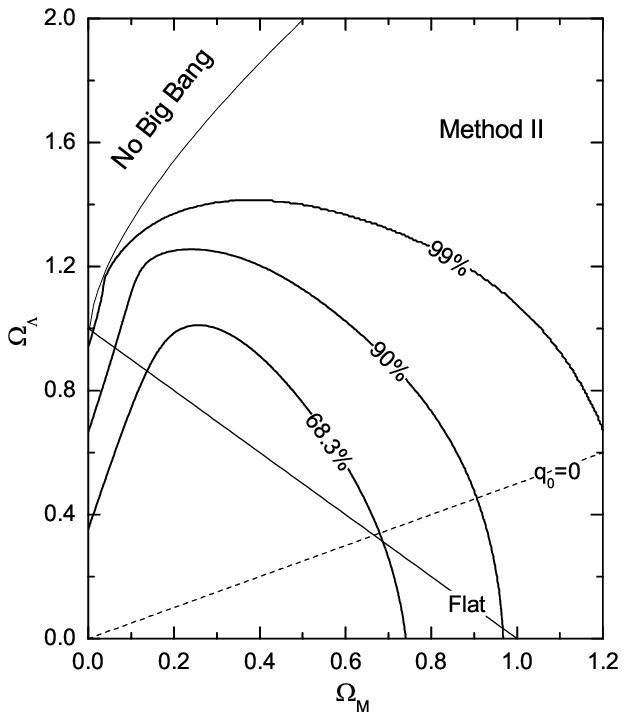} \caption{Joint confidence intervals (68.3\%, 90\% and 99\%) in the
$\Omega_M-\Omega_\Lambda$ plane from the 17 GRBs with Method II.\label{fig3}}
\end{figure}

\begin{figure}
\plotone{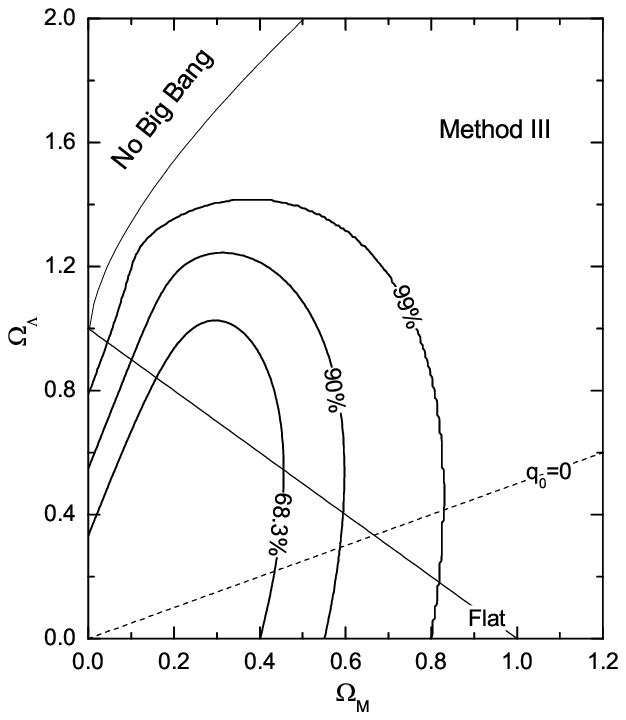} \caption{Joint confidence intervals (68.3\%, 90\% and 99\%) in the
$\Omega_M-\Omega_\Lambda$ plane from the 17 GRBs with Method III. \label{fig4}}
\end{figure}

\begin{figure}
\plotone{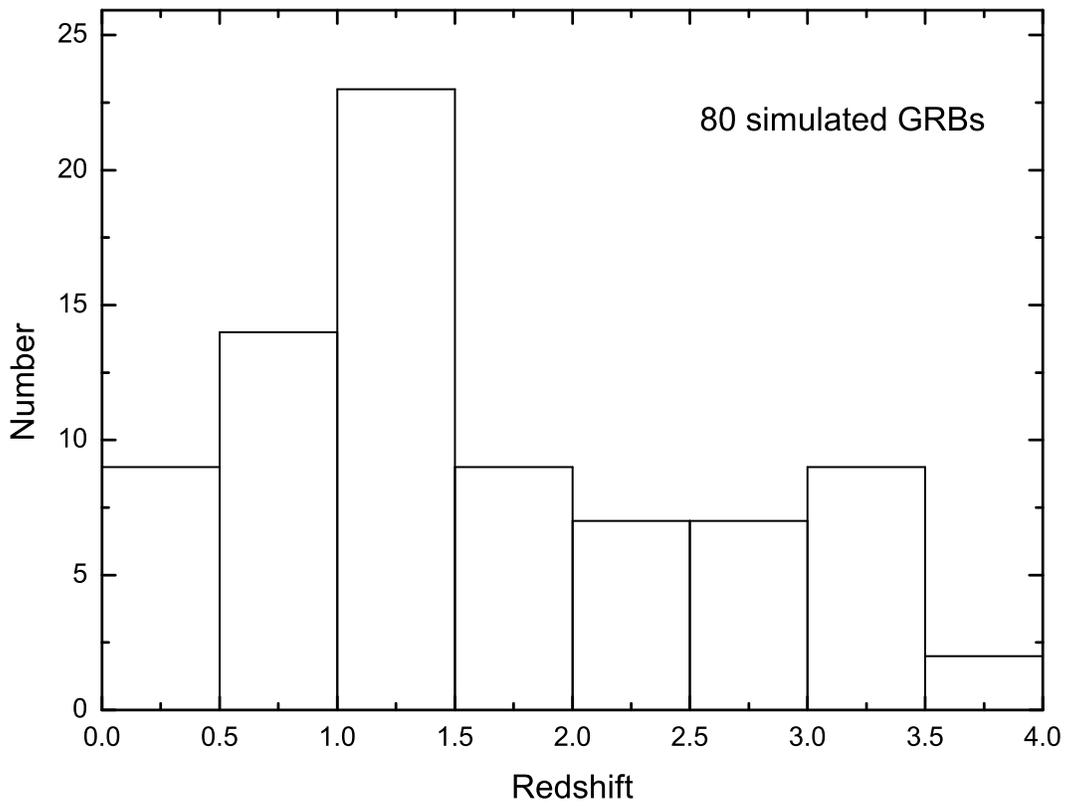} \caption{Histogram for the redshifts of 80 simulated GRBs, following the
redshift distribution of the GRBs observed sofar.\label{fig5}}
\end{figure}

\begin{figure}
\plotone{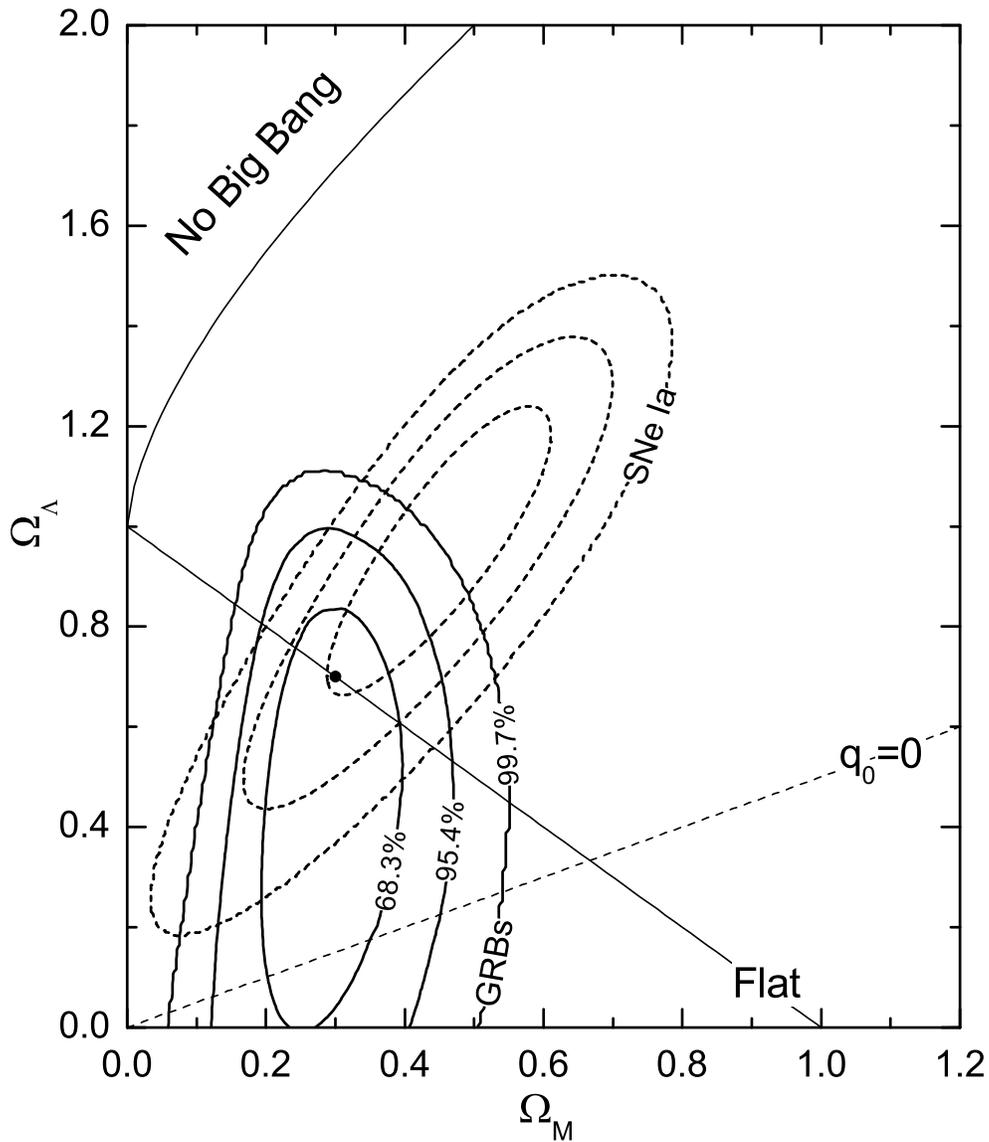} \caption{Joint confidence intervals (68.3\%, 95.4\% and 99.7\%) in the
$\Omega_M$-$\Omega_\Lambda$ plane from the 80 simulated GRBs in this work (solid contours) and
from the 157 SNe Ia in Riess et al. (2004) (dashed contours). The black dot marks
$\Omega_M=0.30$ and $\Omega_\Lambda=0.70$.\label{fig6}}

\end{figure}
\end{document}